\documentstyle[twocolumn,aps,epsf]{revtex}
\begin{document}
%%\draft
\twocolumn[\hsize\textwidth\columnwidth\hsize\csname @twocolumnfalse\endcsname
\title{Comment on ``Hidden assumptions in decoherence theory"}
\author{Tabish Qureshi}
\address{Department of Physics, Jamia Millia Islamia, 
New Delhi-110025, INDIA.\\Email: tabish@arbornet.org}

\maketitle
\begin{abstract}
It is shown that the conclusion of the paper ``Hidden assumptions in
decoherence theory" \cite{vecchi} is the result of a misunderstanding of
the concept of pointer states. It is argued that pointer states are
{\em selected} by the interaction of quantum systems with the environment,
and are not based on any {\em measurement} by a conscious observer.

\end{abstract}
\pacs{PACS nos:}
]

Italo Vecchi has very recently written an article which apparently 
points out some ``hidden assumptions'' in the formulation of the idea
of decoherence \cite{vecchi}. The author claims that there is an 
ambiguity in the formulation of decoherence in that any vector can be 
chosen as a pointer basis. We show here that it is not so, and the 
claim is born out of a misunderstanding of the idea of pointer states.

The author starts his argument by considering a system $S$ which is in
a superposition of certain eigenstates $|n\rangle$ and its environment
$W$ which is in a state $|\Phi_0\rangle$.  The evolution can then be
described as

$$ \sum_nc_n|n\rangle |\Phi_0\rangle \rightarrow exp(iHt)\sum_nc_n|n\rangle 
|\Phi_0\rangle \rightarrow \sum_nc_n|n\rangle |\Phi_n(t)\rangle $$

Now, the author of \cite{vecchi} claims that $|\Phi_n(t)\rangle$ are
pointer states. He further goes on to say that ``any act of measurement on
$W$ induces a collapse of its vectors into one of the pointer vectors". As
we started from the assumption that $W$ is the environment with which
$S$ is interacting, it is ridiculous to talk of a measurement on the
environment.  Environment is not something which can be controlled or
measured. The source of confusion can probably be traced back to the
article by Joos where he uses the states of the apparatus-environment 
combined \cite{joos}.

Pointer states are emergent states of a quantum system, the {\em
pointer}, or the {\em apparatus}, because of interaction with the
environment.  These states of the apparatus emerge as stable states as
a result of environment induced super-selection.  Superposition of these
states would be destroyed by the environment.

In fact, in the above example, if $S$ is assumed to be the {\em apparatus}
interacting with the rest of the world, that is, an environment $W$,
it can be used to understand the concept of pointer states. Of course,
for a rigorous calculation, one has to consider a specific model of the
environment.  In this case, clearly, the pointer states are $|n\rangle$
because $S$ was in some arbitrary state, which was represented as a
superposition of the states $|n\rangle$, and after interacting with the
environment, these states get entangled with certain environment states
$|\Phi_n(t)\rangle$, which will eventually lead to a loss of coherence
between the different $|n\rangle$ states. The states $|n\rangle$ are
not arbitrarily {\em chosen} by any external measurement, but{\em  emerge}
because of the nature of the interaction and nature of $S$ itself.

There are several examples
available in the literature where pointer states are shown to emerge
from an interaction with the environment\cite{dk,anu}. For a harmonic
oscillator interacting with the environment, coherent states emerge as
pointer states \cite{anu}.  In \cite{anu}, a harmonic oscillator is
assumed to be acting as a pointer for measuring a spin-1/2. This calculation
doesn't rely on any concept of predictability sieve. The superpositions
between the coherent states are destroyed because of interaction with
a model environment which has an infinite number of degrees of freedom.

The author's other statement, "...it is based on the unphysical no-recoil
assumption on th scattering process...", is also baseless, because the
whole argument is not based on the no-recoil assumption, which is just an
approximation to derive a simplified result. Apparent diagonalization of 
a system's density matrix can be demonstrated using exact quantum dynamics 
of a model system coupled to a model environment \cite{dk}.

The author's example of Planck's radiation law can be easily understood 
in the light of the recent results of Paz and Zurek \cite{paz} which
show that for quantum systems very weakly coupled to the environment,
energy eigenstates are the pointer states. Thus one knows that the right
thing to do is to apply the entropy maximization to the discreet energy
spectra, and not to any other basis. Thus one doesn't need any observer
for Planck's law, as claimed in \cite{vecchi}. Rayleigh-Jean's law is 
just the long wavelength limit of
Planck's law. In other words, in the appropriate limit, Planck's law
{\em appears to be} Rayleigh-Jeans law, and one does not need to invoke an 
{\em observer} associated with continuous spectra.

In conclusion, we emphasize that the pointer states are the emergent
stable states of a quantum system because of its interaction with the
environment, and are {\em not} an outcome of any measurement by an observer.
Thus there is no hidden assumption in the decoherence theory in this regard,
as claimed in \cite{vecchi}.


\begin{thebibliography}{99}

\bibitem{vecchi} {\em Hidden assumptions in decoherence theory}, I. 
Vecchi, quant-ph/0001021.

\bibitem{joos} {\em Decoherence through interaction with environment},
A. Joos in {\em Decoherence and the appearance of a classical world in
quantum theory}, eds. Guilini et al. (Springer) (1996)

\bibitem{dk} {Environment-induced decoherence: The Stern-Gerlach 
measurement}, A. Venugopalan, D. Kumar, R. Ghosh, Physica {\bf 220} 
(1995) 563.

\bibitem{anu} {\em Pointer states via Decoherence in a Quantum Measurement}, A. 
Venugopalan, quant-ph/9909005 (To appear in Phys. Rev. A)

\bibitem{paz} {\em Quantum limits of decoherence: environment induced 
super-selection of energy eigenstates}, J. P. Paz and W. H. Zurek, 
Phys. Rev. Lett. {\bf 82} (1999) 5181
\end{thebibliography}
\end{document}